# Shape Bifurcation of a Spherical Dielectric Elastomer Balloon under the Actions of Internal Pressure and Electric Voltage

Xudong Liang[1], Shengqiang Cai[1] *

[1] *Department of Mechanical and Aerospace Engineering, University of California, San Diego, La Jolla, CA 92093, USA*

## Abstract

Under the actions of internal pressure and electric voltage, a spherical dielectric elastomer balloon usually keeps a sphere during its deformation, which has also been assumed in many previous studies. In this article, using linear perturbation analysis, we demonstrate that a spherical dielectric elastomer balloon may bifurcate to a non-spherical shape under certain electromechanical loading conditions. We also show that with a non-spherical shape, the dielectric elastomer balloon may have highly inhomogeneous electric field and stress/stretch distributions, which can lead to the failure of the system. In addition, we conduct stability analysis of the dielectric elastomer balloon in different equilibrium configurations by evaluating its second variation of free energy under arbitrary perturbations. Our analyses indicate that under pressure-control and voltage-control mode, non-spherical deformation of the dielectric elastomer balloon is energetically unstable. However, under charge-control or ideal gas mass-control mode, non-spherical deformation of the balloon is energetically stable.

* Correspondent author

E-mail: s3cai@ucsd.edu (Shengqiang Cai)

# 1. Introduction

A soft dielectric membrane can deform by mechanical stretching or applying electric voltage across its thickness. Experiments are abundant showing the interplay between electric field and mechanics in dielectric elastomers [1-4]. For instance, voltage-induced deformation in a free standing dielectric elastomer membrane can hardly exceed 40% due to electromechanical pull-in instability [5], while a prestretched dielectric elastomer membrane or the membrane subjected to a dead load can deform as large as several hundred percent by voltage without failure [6, 7].

Due to the electromechanical coupling, high energy density, easy fabrication and relatively low cost, dielectric elastomers have been recently explored intensively in diverse applications, including artificial muscles [8-11], haptic devices [12, 13], micro-pumps [14-17] and adaptive lens[18-21] to name a few. Among all dielectric elastomer devices, spherical balloon is one of the most frequently used geometries. For example, dielectric elastomer balloons have been proposed to make reciprocating or peristaltic pumps by Goulbourne [15, 16]. Dielectric elastomer balloons have also been developed into tactile devices [4] and spherical actuators and generators.

The wide applications of dielectric elastomer balloon have motivated recent studies of their deformation under different electromechanical loading conditions. Zhu et al. [2] formulated nonlinear vibrations of a spherical dielectric elastomer balloon subjected to a constant internal pressure and an AC voltage. Rudykh et al. [22] predicted snap-through actuation of a thick-walled dielectric elastomer balloon. Li et al. [7] successfully harnessed electromechanical instabilities of a dielectric elastomer balloon to achieve giant voltage-induced expansion of area.

While the deformation of a dielectric elastomer balloon subjected to a voltage and internal pressure has been intensively studied, in most previous studies, spherical deformation is assumed if the dielectric elastomer balloon is initially a sphere. Little efforts have been made, if any, in studying possible non-spherical deformation in a spherical dielectric elastomer balloon subjected to electromechanical loading. However, on the other hand, non-spherical shape bifurcation has been observed in experiments and predicted in theories for a spherical elastomer balloon only subjected to internal pressure. For example, Alexander [23] has reported the observation of non-spherical deformation mode in a neoprene spherical balloon in the inflation process. Linear perturbation analyses, conducted by different researchers [24-26], predicted the existence of non-spherical deformation mode in a spherical elastomer balloon subjected to internal pressure. Moreover, Fu et al. [27] have recently conducted stability analyses on the non-spherical deformation mode and shown that in certain loading conditions, the non-spherical configuration of the balloon can be stable.

Additionally, in the experiments conducted by Li et al. [7], a region on the top of the dielectric elastomer balloon bulged out significantly when the voltage was high. This phenomenon cannot be predicted by their theoretical model. This experimental observation, combined with the previous studies of the elastomer balloon only subjected to internal pressure, indicates the possible shape bifurcation of dielectric elastomer balloon subjected to a combination of internal pressure and electric voltage. In this article, we study the shape bifurcation in a spherical dielectric elastomer balloon subjected to internal pressure and electric voltage. We will also conduct stability analyses for different modes of deformation under different electromechanical loading conditions.

The paper is organized as follows. Section 2 derives the field equations of a spherical balloon subject to internal pressure and electric voltage. Section 3 describes the homogeneous deformation solution of the balloon. We conduct linear perturbation analyses in Section 4 and calculate inhomogeneous deformation of the balloon in Section 5. Finally, in section 6, we conduct stability analyses on different deformation mode of the dielectric elastomer balloon.

## 2. Axisymmetric deformation of a spherical dielectric elastomer balloon subjected to internal pressure and electric voltage

We investigate the deformation of a spherical balloon made by a dielectric elastomer under the actions of internal pressure $p$ and electric potential $\varphi$, as shown in Fig. 1a. The radius of the balloon in the undeformed state is assumed to be $R$. We assume the deformation of the balloon is axisymmetric. A Cartesian coordinate $x$-$z$ is introduced, with the origin located at the center of the undeformed balloon, to describe the deformation (Fig. 1b). The coordinates of a material point A in the undeformed state can be written as,

$$X = R\sin\theta, \tag{1}$$

$$Z = -R\cos\theta. \tag{2}$$

After deformation, as shown in Fig.1b, point A moves to A' with the coordinate,

$$x = x(\theta),\ z = z(\theta). \tag{3}$$

Let $\lambda_1$ and $\lambda_2$ denote the principle stretches of the membrane in the latitudinal direction and the longitudinal direction, so we have

$$\lambda_1 = x/X, \tag{4}$$

$$\lambda_2 = \frac{1}{R}\sqrt{\left(\frac{dx}{d\theta}\right)^2 + \left(\frac{dz}{d\theta}\right)^2}. \tag{5}$$

The force balance in the z direction and the direction normal to the z axis of the balloon can be written as,

$$2S_2 H \sin\theta \frac{dz}{d\theta} = p\lambda_2 x^2, \tag{6}$$

$$S_1 \frac{dx}{d\theta} = R\lambda_2 \frac{d}{d\theta}(S_2 \sin\theta), \tag{7}$$

where $S_1$ and $S_2$ are the nominal stresses in latitudinal direction and longitudinal direction, $H$ is the thickness of the balloon in undeformed state, which is a constant. Using the definition of $\lambda_1$ and $\lambda_2$, and the geometrical relationship $dx/d\theta = R\lambda_2 \sin\alpha$ and $dz/d\theta = R\lambda_2 \cos\alpha$, where $\alpha$ is the angle between the tangential direction of the deformed balloon and the z axis (Fig. 1b), the force balance equations (6) and (7) can be rewritten as,

$$\frac{d\lambda_1}{d\theta} = \lambda_2 \frac{\sin\alpha}{\sin\theta} - \lambda_1 \cot\theta, \tag{8}$$

$$\frac{d\lambda_2}{d\theta} = \left(S_1 - \lambda_2 \frac{\partial S_2}{\partial \lambda_1}\right)\left(\frac{\partial S_2}{\partial \lambda_2}\right)^{-1} \frac{\sin\alpha}{\sin\theta} - \left(S_2 - \lambda_1 \frac{\partial S_2}{\partial \lambda_1}\right)\left(\frac{\partial S_2}{\partial \lambda_2}\right)^{-1} \cot\theta, \tag{9}$$

$$\frac{d\alpha}{d\theta} = \frac{S_1 \cos\alpha}{S_2 \sin\theta} - \frac{p\lambda_1\lambda_2 R}{S_2 H}. \tag{10}$$

The elastomer is assumed to be incompressible, namely,

$$\lambda_1\lambda_2\lambda_3 = 1, \tag{11}$$

where $\lambda_3$ is the stretch in the thickness direction of the membrane.

Constitutive model of ideal dielectric elastomer is adopted here to describe the electromechanical behaviors of the balloon membrane [5]. The electric field $E$ and the electric displacement $D$ is related by the linear equation,

$$D = \varepsilon E, \tag{12}$$

where $\varepsilon$ is permittivity of the elastomer, independent of the deformation and electric field. The electric field in the membrane can be calculated by $E=\varphi/h$, where $\varphi$ is the electric potential difference between the two surfaces of the membrane and $h$ is the thickness of the membrane in the deformed state which may vary from point to point. The electric displacement is equal to the charge density, namely, $D=dQ/da$, where $da$ is the area of an element of the membrane in deformed state and $dQ$ is the amount of charge on each side of the element.

The relation between the nominal stresses and the stretches are,

$$S_1 = \frac{\partial W_s(\lambda_1, \lambda_2)}{\partial \lambda_1} - \frac{\varepsilon E^2}{\lambda_1}, \tag{13}$$

$$S_2 = \frac{\partial W_s(\lambda_1, \lambda_2)}{\partial \lambda_2} - \frac{\varepsilon E^2}{\lambda_2}, \tag{14}$$

where the first terms in both equations are elastic stress and the second terms are Maxwell stress. $W_s(\lambda_1,\lambda_2)$ is the stretching free energy of the elastomer, for which we adopt Ogden model [28],

$$W_s(\lambda_1,\lambda_2) = \sum_{r=1}^{3} \frac{\mu\mu_r}{\alpha_r}(\lambda_1^{\alpha_r} + \lambda_2^{\alpha_r} + (\lambda_1\lambda_2)^{-\alpha_r} - 3), \tag{15}$$

where $\mu$ is the shear modulus for infinitesimal deformation, $\alpha_r$ and $\mu_r$ are the material constants. In this article, we use the following material parameters: $\alpha_1$=1.3, $\alpha_2$=5.0, $\alpha_3$=-2.0 and $\mu_1$=1.491, $\mu_2$=0.003, $\mu_3$=-0.023. Inserting Eq. (15) into Eqs. (13) and (14), we obtain that

$$S_1 = \sum_{r=1}^{3} \mu\mu_r \left( \lambda_1^{\alpha_r-1} - \frac{1}{\lambda_1^{\alpha_r+1}\lambda_2^{\alpha_r}} \right) - \frac{\varepsilon E^2}{\lambda_1}, \tag{16}$$

$$S_2 = \sum_{r=1}^{3} \mu\mu_r \left( \lambda_2^{\alpha_r-1} - \frac{1}{\lambda_1^{\alpha_r}\lambda_2^{\alpha_r+1}} \right) - \frac{\varepsilon E^2}{\lambda_2}. \tag{17}$$

In the southern and northern poles of the balloon, we have the following boundary conditions:

$$\lambda_1(0) = \lambda_2(0), \tag{18}$$

$$\lambda_1(\pi) = \lambda_2(\pi), \tag{19}$$

$$\alpha(0) = \frac{\pi}{2}, \ \alpha(\pi) = -\frac{\pi}{2}. \tag{20}$$

Using constitutive Eqs. (12), (16) and (17), the right hand side of the Eqs. (8)~(10) can be expressed as functions of $\lambda_1(\theta)$, $\lambda_2(\theta)$ and $\alpha(\theta)$. Together with the boundary conditions (18)~(20), the deformation of the dielectric elastomer balloon under different eletromechanical loadings can be calculated.

## 3. Homogeneous deformation

Apparently, homogeneous deformation of the spherical balloon is a solution to the equations in sec.2, i.e,

$$\lambda_1 = \lambda_2 = \lambda_0, \tag{21}$$

$$S_1 = S_2 = S_0, \tag{22}$$

$$E = E_0. \tag{23}$$

The value $\lambda_o$, $S_o$ and $E_o$ depend on the loading conditions, namely, the magnitude of electric potential $\varphi$ and internal pressure $p$. Because the deformation in the balloon is homogeneous, the electric field and stress in the balloon are also homogeneous. As a consequence, a combination of Eqs. (8)~(10) results in a single nonlinear algebra equation for determining stretch $\lambda_o$:

$$\frac{pR}{\mu H}\lambda_0^2 + 2\left(\frac{\varphi}{H\sqrt{\mu/\varepsilon}}\right)^2 \lambda_0^3 - 2\sum_{r=1}^{3}\mu_r(\lambda_0^{\alpha_r-1} - \lambda_0^{-2\alpha_r-1}) = 0, \tag{24}$$

where $pR/\mu H$ is the dimensionless pressure and $\varphi/\left(H\sqrt{\mu/\varepsilon}\right)$ is the dimensionless electric potential. With knowing the homogenous stretch $\lambda_o$, the volume of the balloon can be easily calculated. Fig. 2 plots the volume of the balloon as a function of internal pressure for three different electric voltages. All the three $p$-$V$ curves have a N shape, which are consistent with the results reported previously [25, 27]. Due to the N shaped $p$-$V$ curve, it is known that under pressure control, snap through instability in the balloon can happen when the pressure exceeds the peak value in the $p$-$V$ curve.

## 4. Linear perturbation analysis

With homogeneous deformation, the dielectric elastomer balloon keeps spherical shape. However, as discussed in the introduction, non-spherical deformation in the balloon may also happen. We next conduct linear perturbation analysis to investigate the dielectric elastomer balloon bifurcating from spherical deformation to non-spherical deformation.

Linear perturbation is performed on the state of homogeneous deformation with equal-biaxial stretches $\lambda_o$. The radial and tangential displacement perturbation $\delta_r(\theta)$ and $\delta_t(\theta)$ are assumed to be axisymmetric. So, the coordinates of any material point after perturbation can be written as,

$$x(\theta) = \lambda_0 R \sin\theta + \delta_r(\theta)\sin\theta + \delta_t(\theta)\cos\theta, \qquad (25)$$

$$z(\theta) = -\lambda_0 R \cos\theta - \delta_r(\theta)\cos\theta + \delta_t(\theta)\sin\theta. \qquad (26)$$

Generally speaking, the perturbations of the displacement may result in perturbations of stretches $\delta\lambda_1$ and $\delta\lambda_2$, the nominal stresses $\delta S_1$ and $\delta S_2$ and the internal pressure $\delta p$. Consequently, the force balance equations (6) and (7) can be rewritten as,

$$2H(S_2^0 + \delta S_2)\left(\frac{dz^0}{d\theta} + \frac{d\delta z}{d\theta}\right)\sin\theta = (p^0 + \delta p)(\lambda_2^0 + \delta\lambda_2)(x^0 + \delta x)^2, \qquad (27)$$

$$(S_1^0 + \delta S_1)\left(\frac{dx^0}{d\theta} + \frac{d\delta x}{d\theta}\right) = R(\lambda_2^0 + \delta\lambda_2)\frac{d}{d\theta}\left((S_2^0 + \delta S_2)\sin\theta\right), \qquad (28)$$

The upper index "0" represents the variables in the homogeneous deformation state. All the perturbations in Eqs. (27) and (28) can be expressed by power series of $\delta_r(\theta)$ and $\delta_t(\theta)$. To investigate the critical conditions of the bifurcation, we only keep the linear order terms of $\delta_r(\theta)$ and $\delta_t(\theta)$. Finally, we obtain the following eigenvalue equation of $\delta_r$,

$$\lambda_0 S_1^0 \frac{\partial S_1^0}{\partial \lambda_1}\left((1-t^2)\frac{d^2\delta_r}{dt^2} - 2t\frac{d\delta_r}{dt}\right) + \left(S_1^0 + \lambda_0\frac{\partial S_1^0}{\partial \lambda_1} - \lambda_0\frac{\partial S_1^0}{\partial \lambda_2}\right)\left(2S_1^0 - \lambda_0\frac{\partial S_1^0}{\partial \lambda_1} - \lambda_0\frac{\partial S_1^0}{\partial \lambda_2}\right)$$
$$= C\left(S_1^0 - \lambda_0\frac{\partial S_1^0}{\partial \lambda_1} - \lambda_0\frac{\partial S_1^0}{\partial \lambda_2}\right)t - \frac{\pi R^4 \lambda_0^5}{H}\frac{\partial p}{\partial V}\left(S_1^0 + \lambda_0\frac{\partial S_1^0}{\partial \lambda_1} - \lambda_0\frac{\partial S_1^0}{\partial \lambda_2}\right)\int_{-1}^{1}\delta_r dt, \qquad (29)$$

where $t=\cos\theta$, $C$ is a constant of the integration [26], and $R$ is the radius of the balloon in the reference state. Boundary conditions for $\delta_r$ are $\delta'_r(0)=0$ and $\delta'_r(\pi)=0$. Under pressure-control mode, the last term in Eq. (29) is zero. Eq. (29) is consistent with the equation given in [26, 29], and it is also known as the Legendre's equation and the bounded solution is,

$$\delta_r(t) = DP_n(t) + At + B, \qquad (30)$$

where $A$, $B$ and $D$ are constants, and $P_n(t)$ is the Legendre polynomial of order $n$. For each eigen mode, there is one eigenvalue which corresponds to the critical condition for the bifurcation.

Detailed calculations show that the critical loading conditions for the eigen modes of $\delta_r$ with $n \geq 2$ is physically unrealistic, which is consistent with the conclusion given by Shield et al [30]. So we next only focus on the first two eigen modes in the balloon.

For *n=0*, the eigen mode is a constant which corresponds to a homogeneous perturbation,

$$\delta_r(\theta) = 1,\ \delta_t(\theta) = 0, \tag{31}$$

and the critical condition for the bifurcation is given by,

$$2S_1^0 - \lambda_0 \frac{\partial S_1^0}{\partial \lambda_1} - \lambda_0 \frac{\partial S_1^0}{\partial \lambda_2} + \frac{2\pi R^4 \lambda_0^5}{H} \frac{\partial p}{\partial V} = 0. \tag{32}$$

For *n=1*, the eigen mode represents an inhomogeneous perturbation,

$$\delta_r(\theta) = \cos\theta,\ \delta_t(\theta) = 0, \tag{33}$$

and the critical condition for the bifurcation is given by,

$$S_1^0 - \lambda_0 \frac{\partial S_1^0}{\partial \lambda_1} - \lambda_0 \frac{\partial S_1^0}{\partial \lambda_2} = 0. \tag{34}$$

A combination of Eqs. (16), (17) and (32) or (34) gives a nonlinear algebra equation with single unknown $\lambda_0$, which corresponds to the critical conditions of bifurcation for the mode of *n=0* or *n=1*.

The bifurcation modes for *n=0* and *n=1* are both plotted in Fig. 3. Following literature, we name *n=0* as spherical bifurcation mode and *n=1* as pear-shaped bifurcation mode. The critical conditions for the bifurcation are also calculated and plotted in Fig. 2. The circle and square dots represent the critical conditions for the spherical and pear-shaped bifurcation mode, respectively. As expected, for the spherical bifurcation mode, the critical conditions coincide with the extreme points in the *p-V* curve of the balloon with homogenous deformation.

## 5. Inhomogeneous deformation

The bifurcation analysis conducted in sec. 4 predicts that inhomogeneous deformation mode can exist in a spherical dielectric elastomer balloon subjected to internal pressure and electric voltage. In this section, we conduct post-bifurcation analysis of the dielectric elastomer balloon by numerically solving the governing equations of the balloon formulated in sec.2.

We use shooting method to numerically solve Eqs. (8)~(10). Specifically, the values of the three variables in the southern pole of the balloon are set to be $\lambda_1(0)=\lambda_2(0)=\lambda_a$ and $\alpha(0)=\pi/2$. Those values are used as the initial conditions and Eqs. (8)~(10) can be numerically integrated to obtain $\lambda_1(\theta)$, $\lambda_2(\theta)$ and $\alpha(\theta)$. We continuously vary the value of $\lambda_a$ until the boundary conditions in the northern pole: $\lambda_1(\pi)=\lambda_2(\pi)$ and $\alpha(\pi)=-\pi/2$ are all satisfied.

With a given pressure and electric voltage, we can obtain the solutions for homogeneous deformation of the balloon, which agree with the solution described by Eq. (24). As expected, in addition to the homogeneous deformation, for a certain range of pressure with different voltages, we can also obtain the solution describing inhomogeneous deformation of the balloon, which is also plotted in Fig.2 by dash curves.

To quantitatively describe Fig.2, we mark several key pressures for three different voltages, which are maximum pressure $p_{max}$, minimum pressure $p_{min}$ obtained from the homogeneous deformation and critical pressures for the pear-shaped bifurcations predicted from the linear perturbation analysis: $p_{cr1}$ and $p_{cr2}$. All the four pressures depend on the magnitude of the voltage. For a given voltage, when $p<p_{min}$ (or $p>p_{min}$), the balloon has one equilibrium solution, corresponding to spherical deformation. For $p_{min}<p<p_{max}$, three equilibrium solutions of spherical deformation can be obtained, and a pear-shaped deformation mode exists if the pressure is between the two critical pressures, namely, $p_{cr2}<p<p_{cr1}$. It is also shown in Fig. 2 that the electric field applied to the dielectric elastomer leads to a lower critical pressure for the bifurcation.

Fig. 2 also shows that as the voltage is increased, the difference of the *p-V* curve between the homogeneous deformation and inhomogeneous deformation increases. The results can be qualitatively understood as follows: when the balloon bifurcates from a spherical shape with homogeneous deformation to a non-spherical shape, the thickness of the balloon membrane becomes inhomogeneous, which results in inhomogeneous electric field since the voltage across the membrane is a constant. The inhomogeneous electric filed will induce inhomogeneous Maxwell stress which in-turn further increases the deviation of the non-spherical bifurcated shape from the spherical shape.

Fig. 4 plots the shapes and electric field of the dielectric elastomer balloon in two adjacent deformation modes marked in Fig.2. When the electric potential is zero, the differences of the volume and the geometry between the spherical and non-spherical modes are almost negligible. As the voltage increases, the volume and the geometrical difference between the two modes become more and more obvious. For the non-spherical deformation mode, the electric field in the balloon membrane is highly inhomogeneous.

Fig.5 plots the electric field, stretch and stress distribution in the dielectric elastomer balloon with the deformation modes as shown in Fig.4 for $\varphi/\left(H\sqrt{\mu/\varepsilon}\right)=0.16$. Compared to the homogeneously deformed state, the concentration factor of the electric field can be as large as 500% and the concentration factor of stresses and stretches can be as large as 200%. The high concentration factor explains the experimental observations by Li et al. [7] that localized bulging out in the dielectric elastomer balloon usually immediately leads to electric breakdown of the dielectric membrane.

## 6. Stability analysis

In the previous sections, we have demonstrated that both spherically and non-spherically deformed dielectric elastomer balloons can be in equilibrium states. However, it is still unclear whether the equilibrium states we obtained are stable or not. In this section, we will conduct stability analysis.

Following the energetic method adopted by different researchers [27, 31, 32], we first derive second variation of the free energy of the dielectric elastomer balloon system. If the second variation of the free energy of an equilibrium state is positive definite, the state is energetically stable. On the other hand, if there is any perturbation which can lead to negative second variation of the free energy of an equilibrium state, the state is regarded as energetically unstable.

It is also known that the stability of a structure depends on its loading method. In this article, we focus on four different ways of applying electromechanical loadings onto the dielectric elastomer balloon. In terms of the mechanical loading, we consider either gradually increasing the internal pressure or the number of ideal gas molecules inside the balloon, for which we call pressure-control mode or ideal gas mass-control mode respectively. In terms of electrical loading, we consider either gradually increasing the voltage across the thickness of the dielectric membrane or the total amount of charge on its surface, for which we call voltage-control mode or charge-control mode respectively. Consequently, we have four different combinations of electomechanical loading method. We next derive the second variation of the free energy for the four different cases.

In the pressure-control and voltage-control mode, the balloon, together with the pressure and electric voltage forms a thermodynamic system with the free energy given by,

$$F = \int (W_s(\lambda_1, \lambda_2) + W_e(\lambda_1, \lambda_2))dV_m - pV - \varphi Q, \tag{35}$$

where $W_s(\lambda_1, \lambda_2)$ and $W_e(\lambda_1, \lambda_2)$ are strain energy density and electrostatic energy density of the membrane, $dV_m = 2\pi HR^2 \sin\theta d\theta$ is the volume element of the dielectric membrane, and $V = \int_0^\pi \pi x^2 z' d\theta$ is volume of the balloon. The relationship between electric displacement and electric field in the dielectric membrane is assumed to be linear, the electrostatic energy is,

$$\int W_e(\lambda_1, \lambda_2) dV_m = \int_0^\pi \left(\frac{\varepsilon E^2}{2}\right) 2\pi HR^2 \sin\theta d\theta, \tag{36}$$

where $W_e(\lambda_1, \lambda_2)=\varepsilon E^2/2$ is the electrostatic energy density. Adopting the assumption of ideal dielectric elastomer, $D=\varepsilon E$, and the definition of the electric displacement, $D=dQ/da$, we have,

$$Q = \int_0^\pi (\varepsilon E) 2\pi \lambda_1 \lambda_2 R^2 \sin\theta d\theta, \tag{37}$$

where $da=2\pi\lambda_1\lambda_2 R^2 \sin\theta d\theta$ is the area of the surface element of the membrane in the deformed state. Putting Eqs. (36) and (37) into Eq. (35), with the definition $E=\varphi/h=\lambda_1\lambda_2\varphi/H$, the free energy is in the pressure-control and voltage-control mode,

$$F = 2\pi R^2 H \int_0^\pi \left( W_s(\lambda_1, \lambda_2) - \frac{\varepsilon\varphi^2}{2H^2}(\lambda_1\lambda_2)^2 \right) \sin\theta d\theta - p \int_0^\pi \pi x^2 z' d\theta. \tag{38}$$

Define the first two terms in Eq. (38) as,

$$W(\lambda_1, \lambda_2) = W_s(\lambda_1, \lambda_2) - \frac{\varepsilon\varphi^2}{2H^2}(\lambda_1\lambda_2)^2. \tag{39}$$

For simplicity, $R$ is taken to be unity in the following analysis. Following [27], the second variation of the free energy can be expressed as,

$$\delta^2 F = 2\pi H R^2 \int_0^\pi \left( \mathbf{v}\cdot\mathbf{S}\mathbf{v} + 2\mathbf{v}\cdot\mathbf{L}\mathbf{v}' + \mathbf{v}'\cdot\mathbf{K}\mathbf{v}' \right) d\theta \tag{40}$$

where,

$$\mathbf{S} = \begin{bmatrix} \dfrac{W_{11}}{R^2\sin\theta} - \left(\dfrac{W_{12}}{R^3\lambda_2}x'\right)' - \dfrac{pz'}{HR^2} & 0 \\ 0 & 0 \end{bmatrix} = \begin{bmatrix} \alpha_1 & 0 \\ 0 & 0 \end{bmatrix}, \tag{41}$$

$$\mathbf{L} = \begin{bmatrix} 0 & \dfrac{W_{12}z'}{R^3\lambda_2} - \dfrac{px}{HR^2} \\ 0 & 0 \end{bmatrix} = \begin{bmatrix} 0 & \alpha_2 \\ 0 & 0 \end{bmatrix}, \tag{42}$$

$$\mathbf{K} = \begin{bmatrix} \dfrac{(\lambda_2 W_{22}x'^2 + W_2 z'^2)\sin\theta}{R^4\lambda_2^3} & \dfrac{(\lambda_2 W_{22} - W_2)\sin\theta}{R^4\lambda_2^3}x'z' \\ \dfrac{(\lambda_2 W_{22} - W_2)\sin\theta}{R^4\lambda_2^3}x'z' & \dfrac{(\lambda_2 W_{22}z'^2 + W_2 x'^2)\sin\theta}{R^4\lambda_2^3} \end{bmatrix} = \begin{bmatrix} \alpha_4 & \alpha_5 \\ \alpha_5 & \alpha_3 \end{bmatrix} \tag{43}$$

where $W_1 = \partial W / \partial \lambda_1, W_{12} = \partial^2 W / \partial \lambda_1 \partial \lambda_2$, etc. **v** is a vector of displacement perturbation in the *x* and *z* direction, with **v**=[$\delta x$, $\delta z$]$^T$ and **v'**=[$\delta x'$, $\delta z'$]$^T$.

Next, we derive the second variation of the free energy $\delta^2 F$ of the dielectric elastomer balloon under ideal gas mass-control and voltage-control mode. So, the enclosed ideal gas, the balloon and the electric voltage form a thermodynamic system with the free energy given by,

$$F = \int (W_s(\lambda_1, \lambda_2) + W_e(\lambda_1, \lambda_2))dV_m + \Phi(V, N) - \varphi Q, \tag{44}$$

where $\Phi(V, N)$ is the gas potential energy,

$$\Phi(V, N) = -kTN \ln \frac{V}{V_0}, \tag{45}$$

where *T* is the temperature of the gas, *V* and $V_0$ are the current and initial volume of the gas, *N* is the number of the gas molecules and *k* is the Boltzmann constant. Pressure is defined by $p = -\partial \Phi / \partial V = kTN/V$. The second variation Eq. (44) is,

$$\delta^2 F - \frac{kTN}{V^2}(\delta V)^2 = 2\pi H R^2 \int_0^\pi \left( \mathbf{v} \cdot \mathbf{S} \mathbf{v} + 2\mathbf{v} \cdot \mathbf{L} \mathbf{v}' + \mathbf{v}' \cdot \mathbf{K} \mathbf{v}' \right) d\theta, \tag{46}$$

In the following, we study pressure-control and charge-control mode. When the total amount of charge on the surface of dielectric elastomer membrane is given, the voltage $\varphi$ is an unknown constant. According to Eq. (37) and the definition $E = \lambda_1 \lambda_2 \varphi / H$,

$$E = \frac{Q}{2\varepsilon R^2 \Lambda} \lambda_1 \lambda_2, \tag{47}$$

where $\Lambda = \int_0^\pi \pi(\lambda_1 \lambda_2)^2 \sin\theta d\theta$.

The thermodynamic system under the pressure-control and charge-control mode is formed by the balloon and the pressure, and the free energy of the system is,

$$F = \int (W_s(\lambda_1, \lambda_2) + W_e(\lambda_1, \lambda_2))dV_m - pV. \tag{48}$$

The electrostatic energy of the dielectric membrane is obtained by putting Eq. (47) into (36), and it is expressed as,

$$\int W_e(\lambda_1,\lambda_2)dV_m = \frac{Q^2 H}{4\varepsilon R^2 \Lambda}, \tag{49}$$

The second variation of the free energy is,

$$\delta^2 F - \frac{Q^2 H}{2\varepsilon R^2 \Lambda^3}(\delta\Lambda)^2 = 2\pi HR^2 \int_0^\pi \left(\mathbf{v}\cdot\mathbf{Sv} + 2\mathbf{v}\cdot\mathbf{Lv'} + \mathbf{v'}\cdot\mathbf{Kv'}\right)d\theta, \tag{50}$$

where $\delta\Lambda$ is the first variation of $\Lambda$.

For ideal gas mass-control and charge-control mode, the dielectric membrane balloon and the enclosed ideal gas form a thermodynamic system with the free energy,

$$F = \int (W_s(\lambda_1,\lambda_2) + W_e(\lambda_1,\lambda_2))dV_m - kTN\ln\frac{V}{V_0}. \tag{51}$$

Based on the previous derivations, it is easy to show that the second variation of the free energy,

$$\delta^2 F - \frac{Q^2 H}{2\varepsilon R^2 \Lambda^3}(\delta\Lambda)^2 - \frac{kTN}{V^2}(\delta V)^2 = 2\pi HR^2 \int_0^\pi \left(\mathbf{v}\cdot\mathbf{Sv} + 2\mathbf{v}\cdot\mathbf{Lv'} + \mathbf{v'}\cdot\mathbf{Kv'}\right)d\theta. \tag{52}$$

Without further calculations, by comparing Eqs (40), (46), (50) and (52), we can conclude that for the same equilibrium state, the second variation of the free energy is the smallest for pressure-control and voltage-control mode but the largest for mass-control and charge-control mode, which indicates that pressure-control and voltage-control mode is the least stable while the mass-control and charge-control mode is the most stable.

It can be further proved that second variation of free energy for all four different loading cases can be evaluated by solving the following eigenvalue equation [27]:

$$\mathbf{Sv} + \mathbf{Lv'} - (\mathbf{L}^T\mathbf{v} + \mathbf{Kv'})' - \frac{\delta p x}{HR^2}\begin{pmatrix} z' \\ -x' \end{pmatrix} - \frac{Q\delta\varphi}{HR^6\Lambda}\begin{pmatrix} \frac{x(x'^2+z'^2)}{\sin\theta} - \left(\frac{x^2 x'}{\sin\theta}\right)' \\ -\left(\frac{x^2 z'}{\sin\theta}\right)' \end{pmatrix} = \alpha\sin\theta\,\mathbf{v}, \tag{53}$$

of which eigen mode can be viewed as the perturbation and the eigenvalue $\alpha$ is exactly the second variation of the free energy $\delta^2 F$. The last two terms on the left hand side of Eq. (53) is zero if the balloon is in the pressure-control and voltage-control mode, while $\delta p = -kTN\delta V/V^2$ and $\delta\varphi = -QH\delta\Lambda/2\varepsilon R^2\Lambda^2$ for the mass-control and charge-control mode, respectively.

The eigenvalue problem Eq. (53) can be solved by shooting method with boundary conditions,

$$v_1(0) = v_1(\pi) = 0. \tag{54}$$

In pressure-control and voltage-control mode, we can find at least one negative eigenvalue for non-spherical deformation mode and the spherical deformation mode in the descending path of the $p$-$V$ curve shown in Fig.2. The negative eigenvalue is plotted in Fig. 6 for two different voltages. The calculation indicates that both non-spherical deformation mode and spherical deformation mode in the descending path of $p$-$V$ curve are energetically unstable.

For mass-control or charge-control mode, one additional constraint equation for the perturbations needs to be satisfied, which is

$$\left(\frac{HR^2V}{p\pi} + \int_0^\pi \frac{x^2}{2(\alpha_3-\alpha)}d\theta\right)\frac{\delta p}{HR^2} + \left(\int_0^\pi \frac{x^4 z'}{(\alpha_3-\alpha)\sin\theta}d\theta\right)\frac{Q\delta\varphi}{HR^6\Lambda} = -2\int_0^\pi c_1 v_1 d\theta, \tag{55}$$

$$\left(\int_0^\pi \frac{x^4 z'}{2(\alpha_3-\alpha)\sin\theta}d\theta\right)\frac{\delta p}{HR^2} + \left(\frac{\varepsilon R^{12}\Lambda^3}{\pi Q^2} + \int_0^\pi \frac{x^4 z'^2}{(\alpha_3-\alpha)\sin^2\theta}d\theta\right)\frac{Q\delta\varphi}{HR^6\Lambda} = -\int_0^\pi c_2 v_1 d\theta, \tag{56}$$

where $c_1$ and $c_2$ are

$$c_1 = -\frac{\alpha_2 x^2}{2(\alpha_3-\alpha)} + \frac{1}{2}\left(\frac{\alpha_5 x^2}{\alpha_3-\alpha}\right)' + xz', \tag{57}$$

$$c_2 = -\frac{\alpha_2 x^2 z'}{(\alpha_3-\alpha)\sin\theta} + \left(\frac{\alpha_5 x^2 z'}{(\alpha_3-\alpha)\sin\theta}\right)' + \frac{x(x'^2+z'^2)}{\sin\theta} - \left(\frac{x^2 z'}{\sin\theta}\right)'. \tag{58}$$

Because of the additional constraints, it is much more difficult to find a negative eigenvalue for Eq. (53). Detailed calculations show that for all the other three loading methods, the non-spherical deformation of the dielectric elastomer balloon is energetically stable.

At last, we would like to add one more comment on the mass-control and voltage-control mode. It is well known that under pressure-control mode, snap-through instability in a balloon may happen during its inflation process. Such instability can be eliminated by adopting ideal gas mass-control mode. As shown in Fig.2, if only spherical deformation in the balloon is considered, for all three different electric voltages, only one equilibrium solution exists for the mass-control loading mode, which is the crossing point between the curve representing ideal gas law and the calculated *p-V* curve for the balloon, so no instability will happen. However, if non-spherical deformation is not excluded, even in ideal gas mass-control loading mode, multiple solutions can coexist, namely, there may be several crossing points between the curve representing ideal gas law and the calculated *p-V* curves for the balloon as shown in Fig. 2. To be more explicit, we also plot the mass of idea gas molecules as a function of the balloon volume in Fig. 7 for three different voltages. For a fixed number of idea gas molecules in a certain range, multiple equilibrium solutions which correspond to spherical and non-spherical deformation of the balloon can be found. Moreover, we also found that in the ideal gas mass-control mode, once non-spherical deformation of the balloon is an equilibrium solution, it always has lower free energy than the spherically deformed balloon as shown in the inset of Fig. 7. This indicates that non-spherical deformation is more energetic favorable.

**7 Concluding remark**

This paper studies shape bifurcation of a spherical dielectric elastomer balloon subjected to internal pressure and electric voltage. Using linear perturbation analysis, we obtain the bifurcation mode and the corresponding critical conditions of a dielectric elastomer balloon under the action of internal pressure and electric voltage. By numerically solving the governing equations of the dielectric elastomer balloon with axisymmetric deformation and under different electromechanical loading conditions, we obtain both spherical deformation and non-spherical deformation solutions for the balloon. Our calculations further show that shape difference between two adjacent spherical and non-spherical deformation modes can be greatly enhanced by increasing the electrical voltage. The non-spherical deformation of the dielectric elastomer balloon in-turn induce large electric field concentration and stress/stretch concentration in certain area of the balloon, which may lead to the failure of the system. Finally, we calculate second variation of the free energy of the balloon in different equilibrium states. Our calculations demonstrate that non-spherical deformation of the balloon can be either energetically stable or unstable depending on the eletromechanical loading method.

**Acknowledgement**

Shengqiang Cai acknowledges the startup funds from the Jacobs School of Engineering at UCSD.

**Figure**

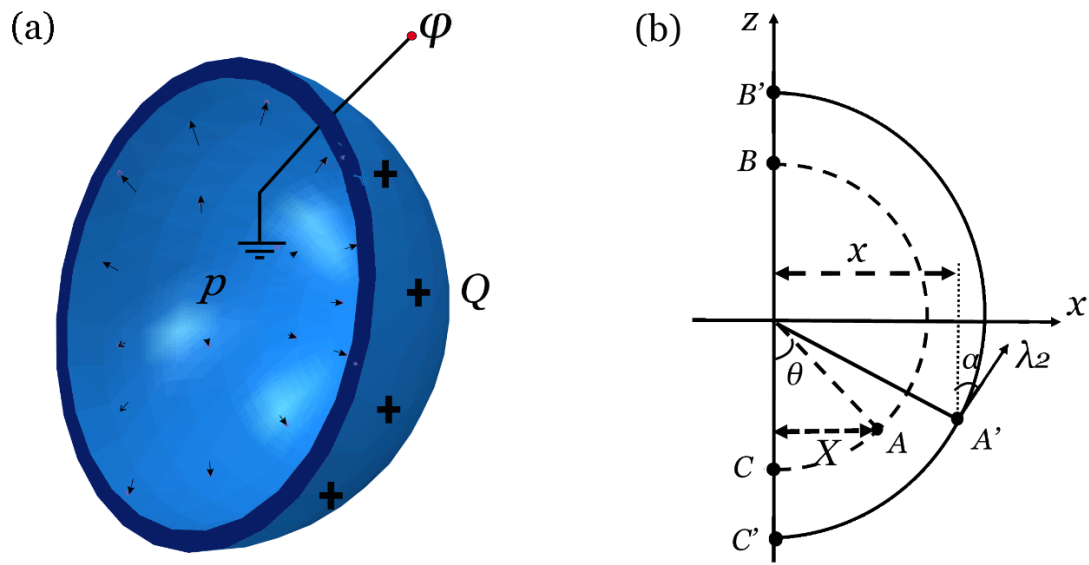

Fig1. (a) Schematics of a dielectric elastomer balloon subjected to internal pressure and electric voltage. (b) The balloon with axisymmetric deformation. Dash line represents the undeformed spherical balloon and solid line represents shape of the balloon after deformation.

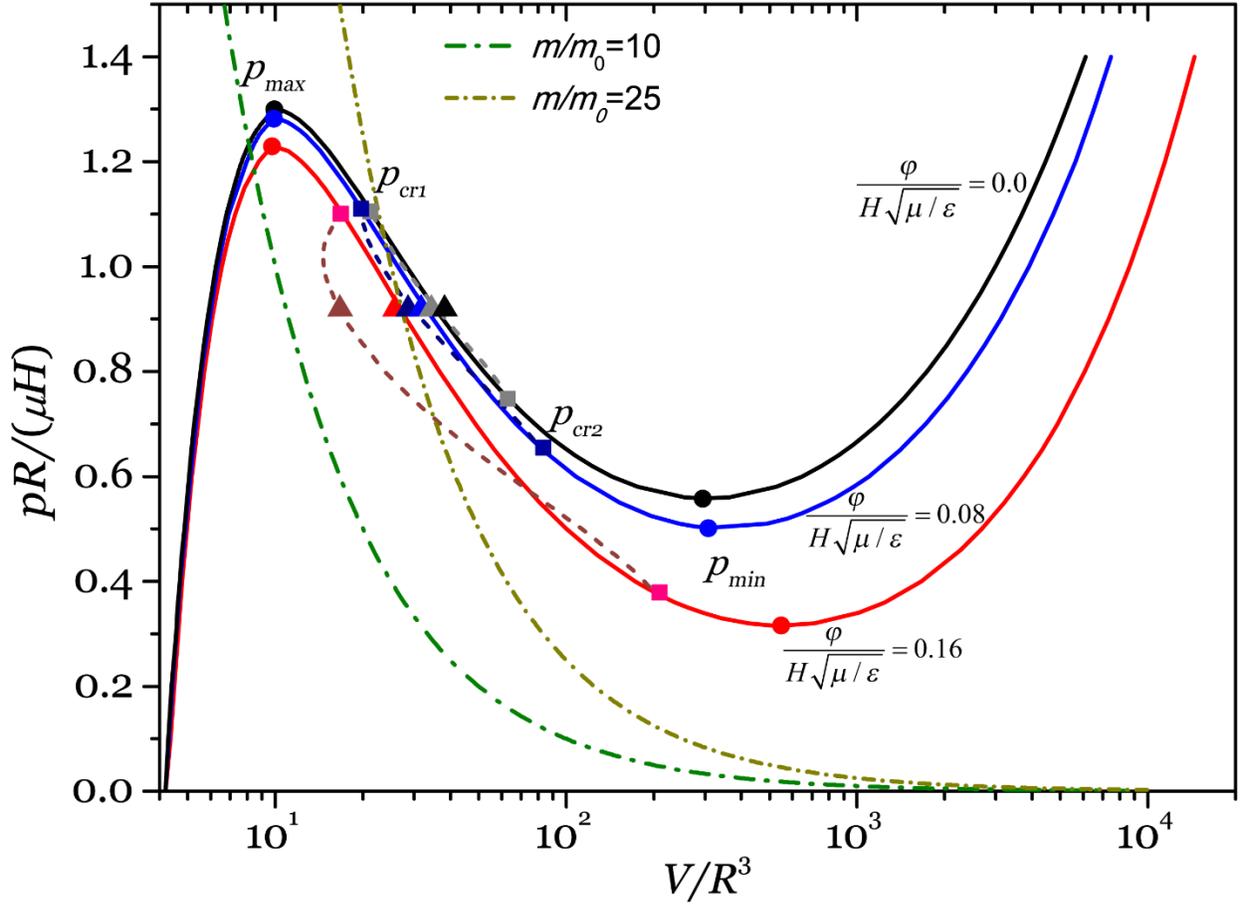

Fig2. The pressure-volume ($p$-$V$) relation of a spherical dielectric elastomer balloon subjected to three different voltages. During the deformation, the balloon may keep a sphere which is represented by the solid curves or become non-spherical which is represented by dash curves. The circle and square dots stand for the bifurcation points predicted from the linear perturbation analysis for spherical and pear-shaped mode respectively. Two adjacent deformation modes with pressure of $pR/(\mu H) = 0.9$ and three different voltages are marked by triangles. The dash-dot lines represent ideal gas law for two different mass of ideal gas, where $m_o$ is the mass of gas molecules when pressure and volume are unity.

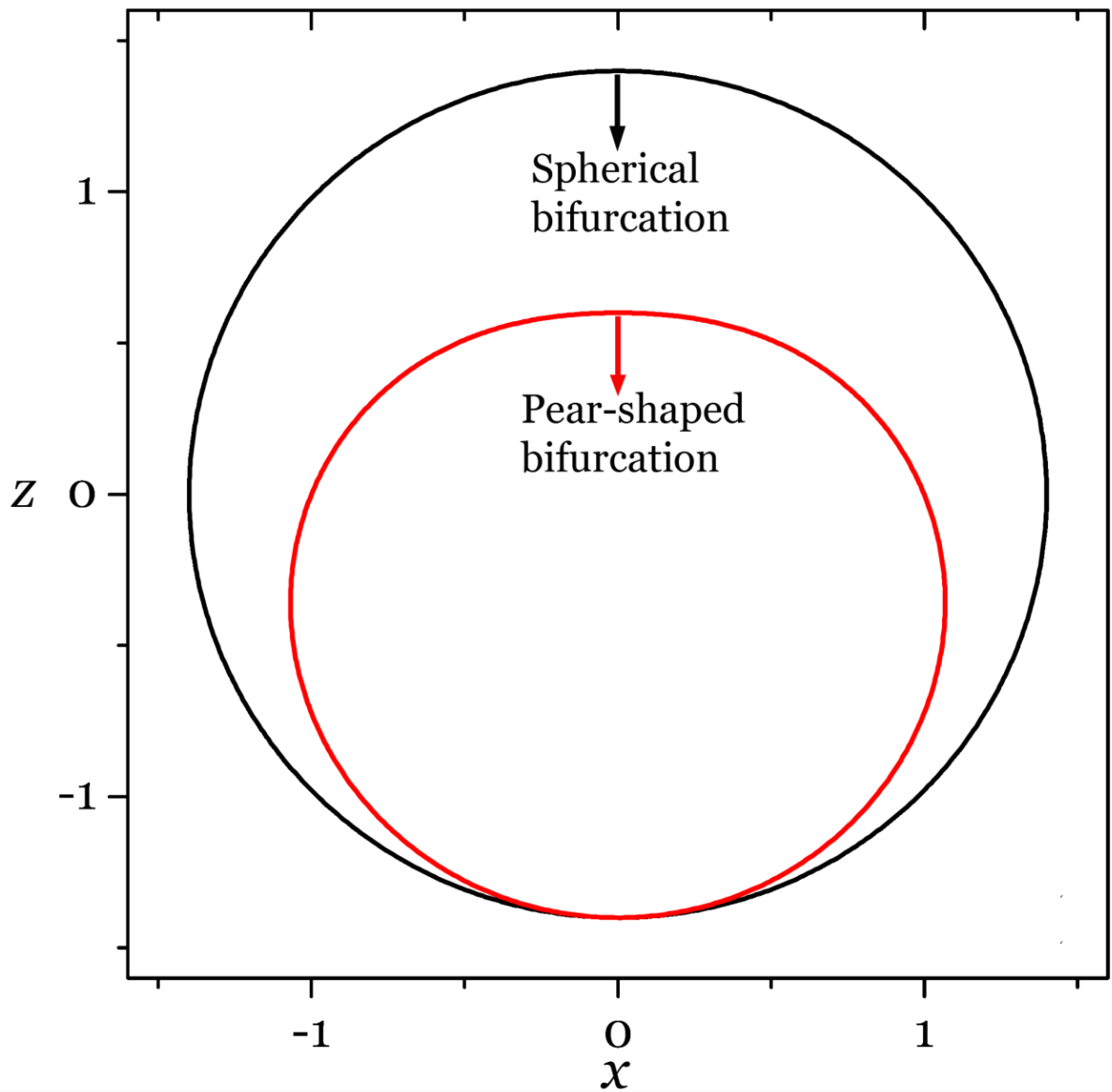

Fig3. Spherical and pear-shaped bifurcation modes calculated from the linear perturbation analysis.

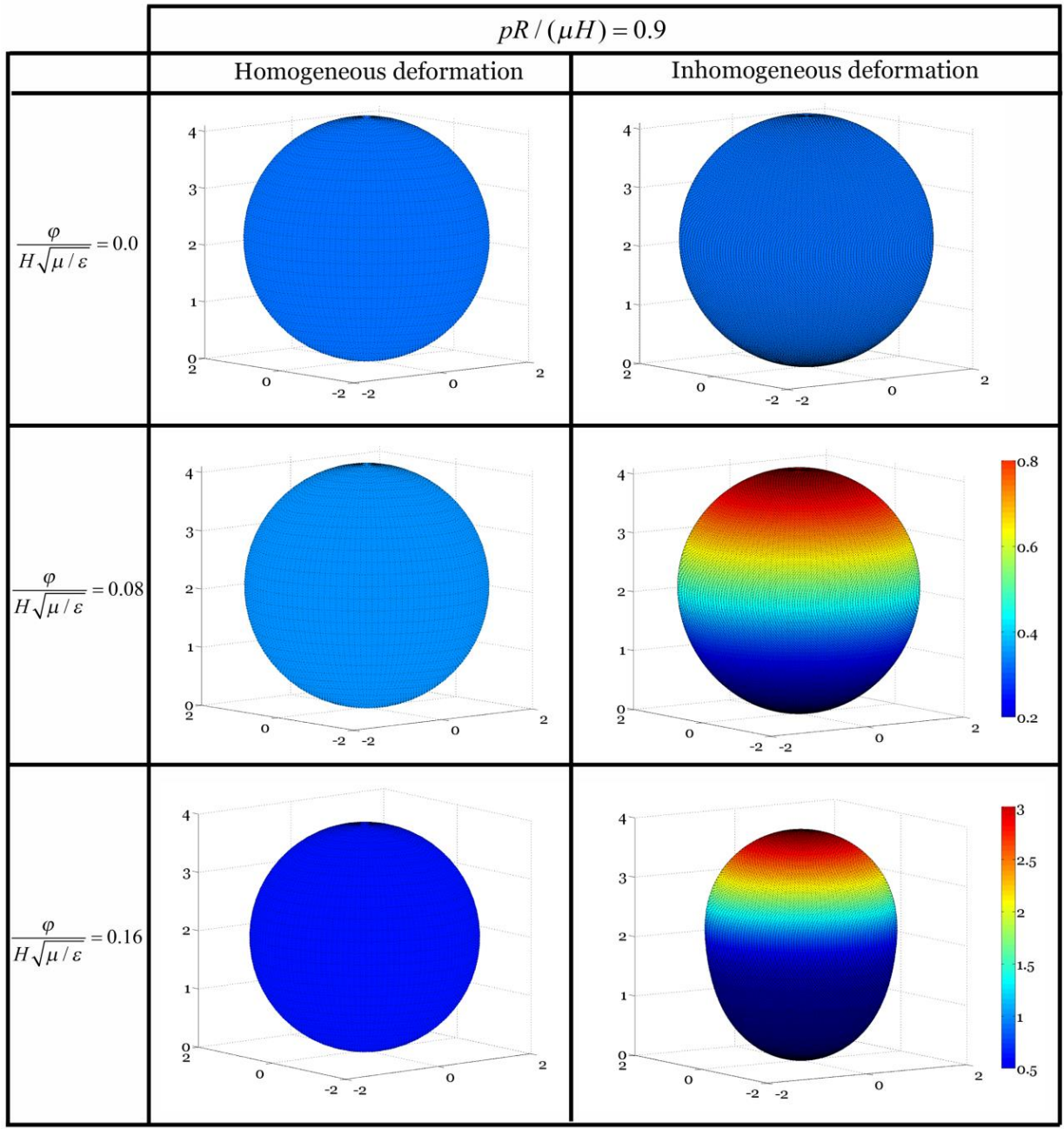

Fig4. Calculated shapes and electric field in a spherical dielectric elastomer balloon in two adjacent deformation modes (spherical mode and pear-shaped mode) marked by triangles in Fig.2. When the electric voltage is high, large electric field concentration can be observed in the pear-shaped mode (right column).

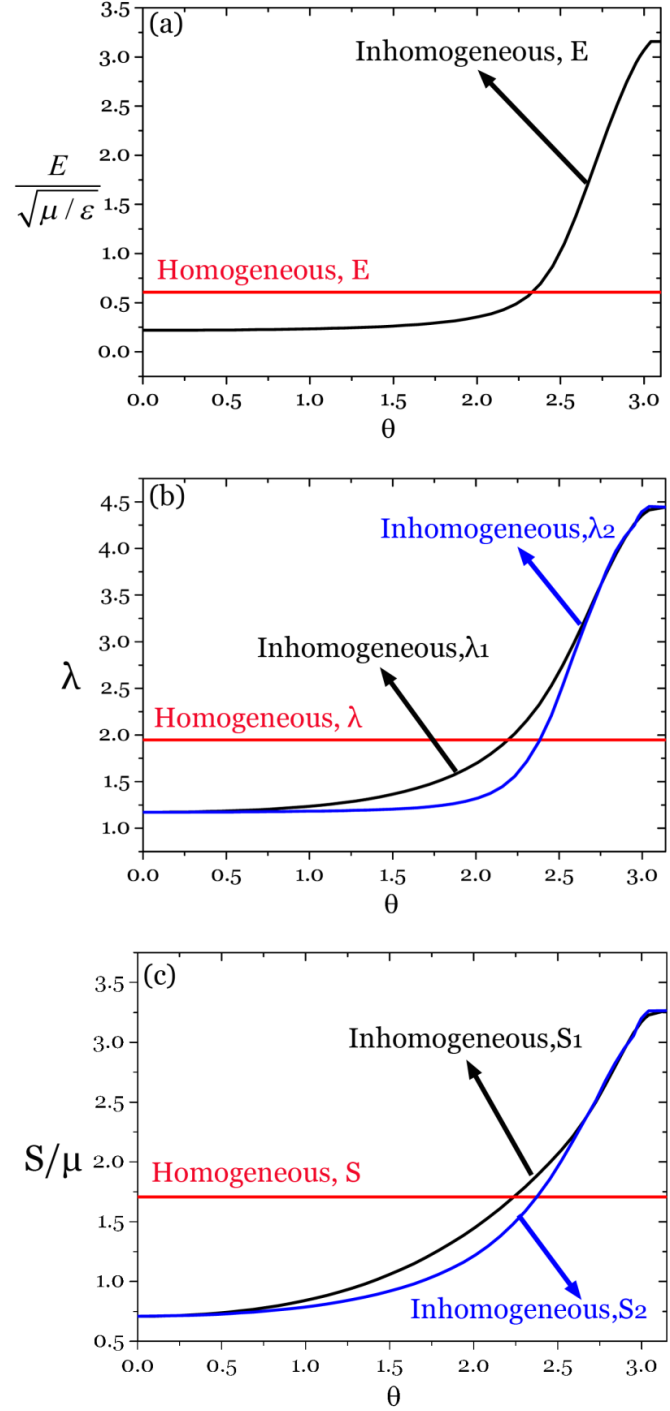

Fig5. Distribution of the electric field, stretch and nominal stress in the dielectric elastomer balloon for homogenous and inhomogeneous deformation modes for $\varphi/\left(H\sqrt{\mu/\varepsilon}\right)=0.16$ as shown in Fig.4.

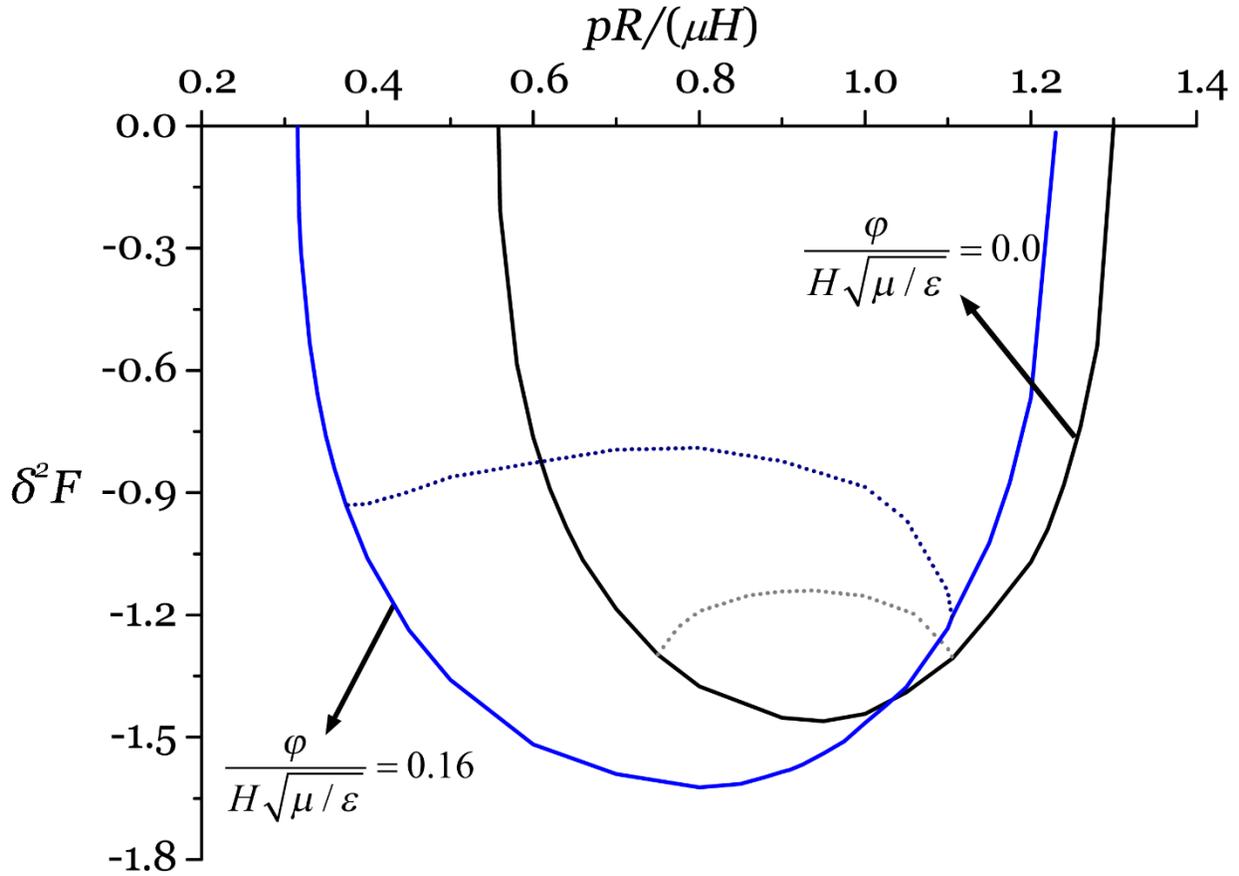

Fig6. Under pressure-control and voltage-control mode, second variation of free energy of the spherical deformation in the descending path of *p-V* curve in Fig.2 and non-spherical deformation can be negative. The solid line represents a negative value for the spherical deformation mode while the dash line shows a negative value for the non-spherical deformation mode. The results indicate that under pressure-control and voltage-control mode, both spherical deformation in the descending path of p-V curve and non-spherical deformation of the dielectric elastomer balloon are energetically unstable.

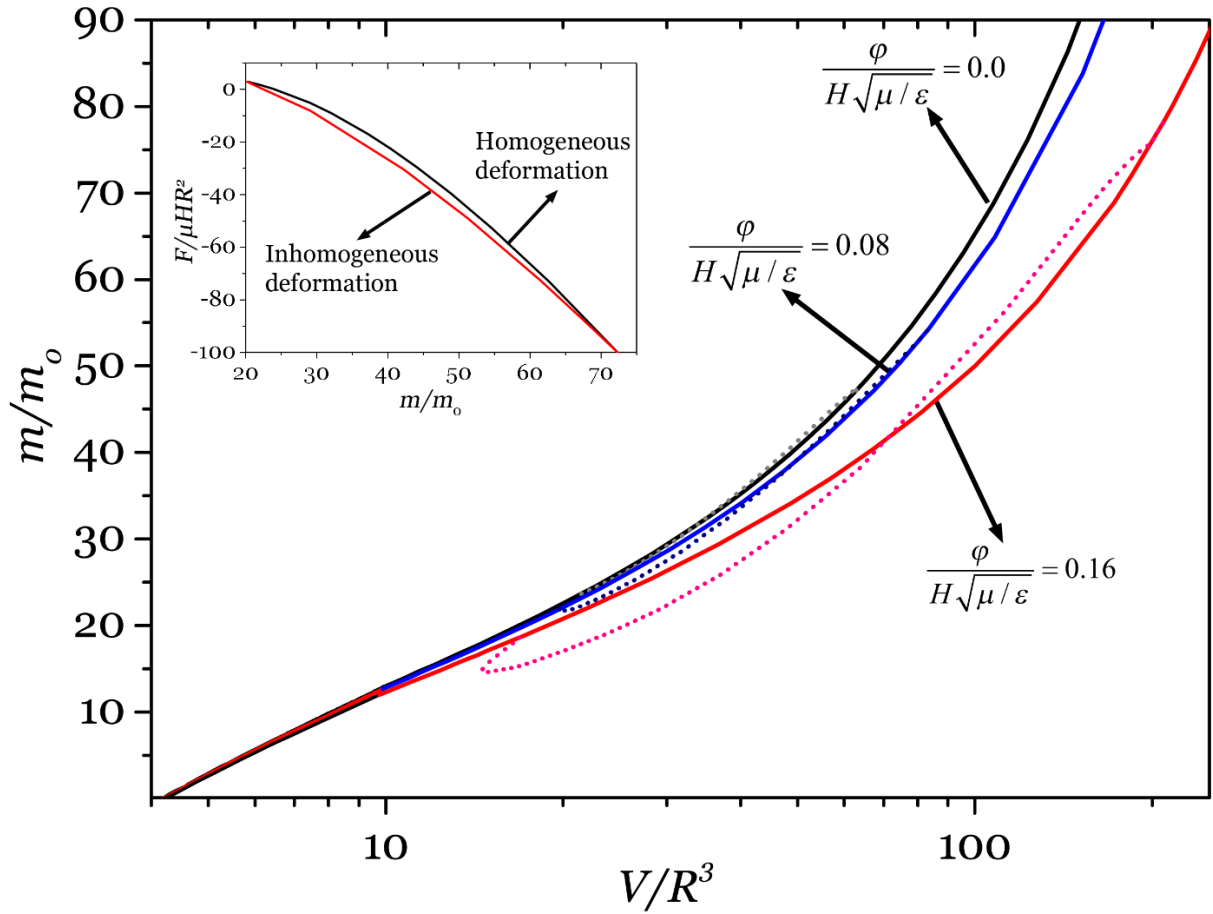

Fig7. The ideal gas mass-volume (*m-V*) curves for a dielectric elastomeric balloon subjected to three different voltages. During the deformation, the gas molecules obey the ideal gas law. The spherical deformation is represented by the solid line and the non-spherical deformation is represented by dash line.